\documentclass[useAMS,usenatbib,referee,usegraphicx]{mn2e}

\newcommand{\kms}{km\,s$^{-1}$}
\newcommand{\percent}{per cent}

\begin{document}
\title[Time Series Spectroscopy of G~29-38]{Deciphering the Pulsations of G~29-38 with Optical Time Series Spectroscopy}   
\author[S. E. Thompson et al.]{Susan E. Thompson$^{1,2}$\thanks{email: sthomp@udel.edu}, M. H. van Kerkwijk$^3$, and J. C. Clemens$^4$\\
$^1$Dept. of Physics and Astronomy, University of Delaware, 217 Sharp Lab, Newark, DE, 19716  USA\\
$^2$Dept. of Physics, Colorado College, Colorado Springs, CO 80903 USA\\
$^3$Dept. of Astronomy and Astrophysics, University of Toronto, 50 Saint George St., Toronto, ON M5S 3H4, Canada\\
$^4$Dept. of Physics and Astronomy, University of North Carolina-Chapel Hill, Chapel Hill, NC 27599 USA\\}

\date{}

\maketitle

\begin{abstract} 
We present optical time series spectroscopy of the pulsating white dwarf star G~29-38 taken at the Very Large Telescope (VLT). By measuring the variations in brightness, Doppler shift, and line shape of each spectrum, we explore the physics of pulsation and measure the spherical degree ($\ell$) of each stellar pulsation mode.  We measure the physical motion of the g-modes correlated with the brightness variations for three of the eight pulsation modes in this data set.  The varying line shape reveals the spherical degree of the pulsations, an important quantity for properly modeling the interior of the star with asteroseismology. Performing fits to the H$\beta$, H$\gamma$, and H$\delta$ lines, we quantify the changing shape of the line and compare them to models and previous time series spectroscopy of G~29-38. These VLT data confirm several $\ell$ identifications and add four new values, including an additional $\ell$=2 and a possible $\ell$=4. In total from both sets of spectroscopy of G~29-38, eleven modes now have known spherical degrees. 
\end{abstract}
\begin{keywords}
stars: oscillations -- stars: white dwarfs -- stars: individual : G~29-38 
\end{keywords}

\section{Introduction}
   
DAVs, or ZZ Cetis, are variable white dwarf stars with hydrogen atmospheres. They are the  coolest known white dwarf pulsators and reside in an instability strip near 12,000~K. Their pulsations are multi-periodic with amplitudes as high as 5~\percent\ and periods between 70~s and 1100~s. In the past few years the number of known DAVs has increased from a few dozen to more than a hundred \cite[see][]{voss07,castanheira06,mullally05,mukadam04}.
The pulsators at the hot end of the instability strip show lower amplitudes and more stable pulsations, while the others show more complex pulsations including variability of the excited modes, and an abundance of harmonic and combination modes.
   
The high gravity (log(g)$\sim$8 cgs) of the DAVs favor gravity-mode type pulsations \citep{rkn82}, where motion along the surface of the star produces regions with varying temperatures.  To describe the spatial distribution of these regions, each of the pulsations on the star are characterized by a spherical harmonic (Y$_{\ell,m}$) and the radial order, $n$. The spherical degree, $\ell$, and azimuthal order, $m$, describe the number of nodal lines across the surface of the star, and directly affect observations of the pulsations.
   
The pulsation's radial component allows asteroseismological models 
to determine the internal structure of the star. The DAVs have evaded unambiguous asteroseismological modeling in part because of the difficulty in measuring the spherical degree of their pulsation modes.  Usually a value of $\ell$=1 is assumed in these models. 
However, it is possible to fit modes with higher values of $\ell$ and this can significantly change the parameters of the model. For example, \citet{bradley06} showed that using the identification of $\ell$=4 \citep{thompson04}, instead of $\ell$=2 \citep{cast04} for one mode, changes the modeled depth of the Hydrogen layer in G~185-32 by a factor of 100.
   
One method to determine $\ell$ independent of asteroseismological modeling was introduced by \citet{robinson95}. They showed that $\ell$ can be obtained from time series optical and UV spectroscopy. The effects of limb darkening increase in the blue wavelengths and decrease through the spectral lines, changing the fraction of integrated light coming from the limb.  As such, the pulsation amplitude varies with both wavelength and the spherical degree of the mode. \citet*{c00} (hereafter C00) measured chromatic amplitudes (pulsation amplitude at each wavelength bin) across the Hydrogen Balmer lines and demonstrated that spectral variations can indeed be used to identify spherical degree. The same method was applied to the DAVs HS~0507+0434B, G~117-B15A, and G~185-32 \citep{kotak02a, kotak04, thompson04} and the DBV GD~358 \citep{kotak03}.  Generally, the method worked well, although for
one mode -- the 272~s mode in G~117-B15A -- the chromatic amplitudes
are extremely puzzling \citep{kotak04}.  Also, they found that the chromatic amplitudes provided unambiguous identifications only for the strongest modes; for weaker ones, noise obscured the expected visible differences.  In order to improve the precision, \citet{kotak03} attempted to use the slope and curvature in selected wavelength ranges of the chromatic amplitudes as measures of spherical degree, with somewhat limited success.  A different approach was demonstrated by \citet{thompson04} where they analysed fits to the spectral lines rather than the observed fluxes in each spectral bin. The chromatic amplitudes created from these fits did allow for mode identifications from lower signal-to-noise data.  
   
G~29-38 is one of the brightest DAVs with complex pulsations and some of the largest amplitudes, making it a prime candidate for time series spectroscopy. It has an abundance of modes, but they are not all excited simultaneously and many observation runs are necessary to measure them all.  \citet{kleinman98} compiled 10~years of photometric observations and found a possible series of $\ell$=1 modes on G~29-38.  Time series spectroscopy taken at the Keck Observatory with the LRIS spectrograph provided chromatic amplitudes of G~29-38 that directly detected an $\ell$=2 mode at 776~s (C00) and gave strong evidence for another at 920~s \citep{kotak02a}. Using the same data, \citet*{vk00} (hereafter VK00) measured the periodic line-of-sight velocities associated with the longitudinal motion of the gas. For the 776~s mode they found relatively large motion as compared to the measured flux amplitude, again indicating that the mode is $\ell$=2 \citep{dz82}.  

The measurements of the pulsation velocities were confirmed by \citet{thompson03} using a series of high resolution spectra of G~29-38 and provided observational evidence regarding the convective driving theory proposed for DAVs \citep{bhill83, gw1}.  The velocities also hint at the nature of the many combination and harmonic modes present on these pulsators. These combinations and harmonics most likely result from nonlinear mixing at the surface of the star and are not real modes that probe the interior \citep{wu01, montgomery05}. VK00 reinforced this idea by finding no velocity signature associated with the combinations and harmonics on G~29-38. However, \citet{thompson03} reported one combination mode with a significant velocity detection.

In this paper we show the results of time series spectroscopy of G~29-38 taken at the VLT (Very Large Telescope) in 1999.  We present the light and velocity curves measured from the series of spectra.  We fit 13 modes in the light curve and find variations at the same period in the velocity curve for the largest amplitude pulsations. We then perform fits to the Balmer lines to quantify the changing line shape.  Since we are using fits to the observed spectral lines, we can present the line shape variations in terms of the fitted values: the normalized equivalent width variations of the Gaussian and Lorentzian functions. We perform the same analysis on the 1996 Keck spectral series presented by VK00 and C00 as a check of our method and recover their results of an $\ell$=2 mode among a series of $\ell$=1 modes. Application of the same method to our VLT data adds four additional mode identifications to the G~29-38 pulsation spectrum.

\section{Observations}
We present a 6.14 hour spectral series (604 spectral images) of G29-38 from the VLT, observed with the FORS1 spectrograph \citep{fors} on August 27, 1999 starting at 03:00:19.67 UT. The average exposure time on each image is 16.0~s with a readout time of 16.0~s. We perform a standard reduction on each image using routines in IRAF. We subtract the bias measured from the overscan, divide by a normalized average flat taken at the beginning of the night, and perform wavelength calibration from arc lamps taken at the end of the night. The average flat field was insufficient to remove all the effects of erroneous pixels, thus small dips and peaks appear in the average spectrum. Overall, these are smaller than the average noise in the individual spectrum. Also since we rely on relative fluxes, poor flat-fielding does not significantly change our results. No spectro-photometric standard was available to flux calibrate the spectra. Each extracted spectrum spans 3300~\AA\ to 5700~\AA, has a dispersion of 2.35~\AA\ per binned pixel, and a signal-to-noise of 80. Because light entered through a wide slit, the resolution is set by the seeing, yielding $\sim$8~\AA. 

Figure~1 depicts an example of an individual spectrum and the average of the 604 spectral images. Besides the Hydrogen lines ranging from H$\beta$ to H$9$, the Ca~K line is evident at 3933~\AA. \citet{vonHippel07a} use all known calcium line equivalent width measurements, including those presented here and in VK00, to show that the abundance of Ca on G~29-38 varies by as much as 70 \percent, though \citet{debes08} has since seen no variations in the Ca of G~29-38. Measurements of changing Ca abundance will allow us to constrain the diffusion rate of calcium in the WD atmosphere and the variable accretion from the debris disk surrounding G~29-38 \citep{zuckerman87, reach05}. 

\begin{figure}
\includegraphics[scale=.61, angle=270]{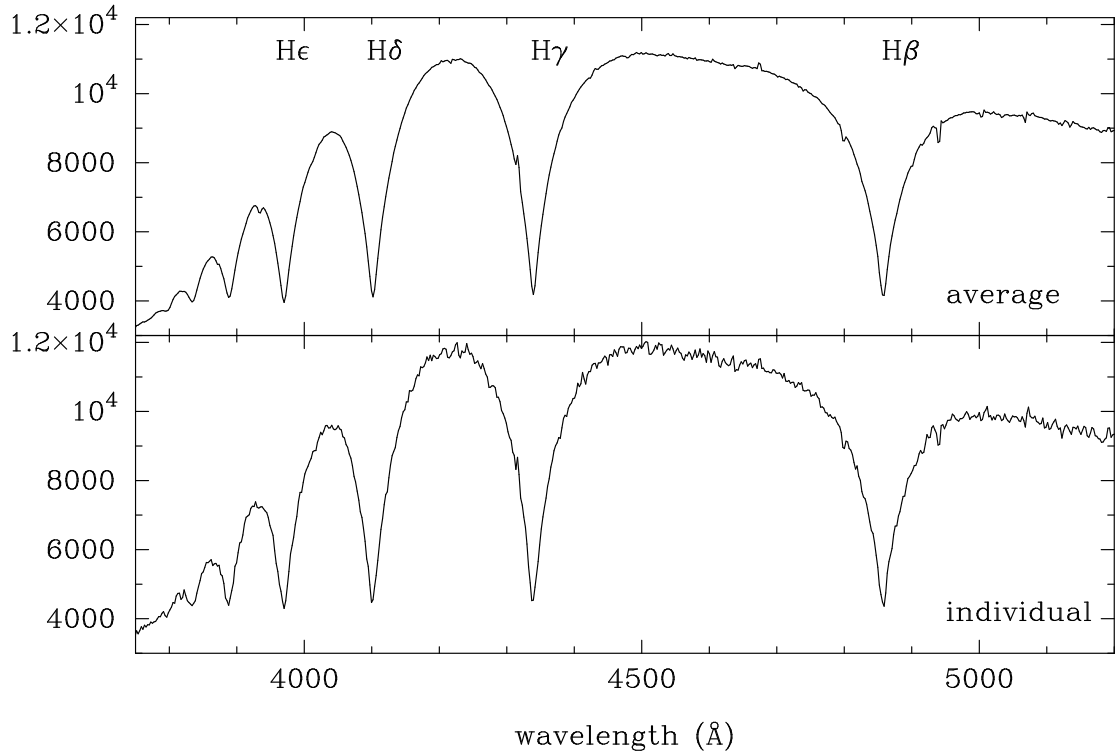}
\caption{Average and individual spectra taken with the FORS1 spectrograph. All small dips in the average spectrum are a result of poor flat-fielding, except for the Ca II line at 3933\,\AA and the Mg I line at 4481\,\AA.}
\end{figure}

\section{Periodicities}
Time series spectroscopy of pulsating white dwarfs provides the opportunity to measure both the variations in brightness of the star and the radial velocity of the gas along the surface of the star. We find periodic variations in our measurements of both the light and velocity curves. We create a light curve by averaging the counts collected in the continuum red-ward of the H$\beta$ line (5000-5647~\AA). The average noise level of the light curve is 0.89 mma (0.089 \percent), measured from the high frequency end of the Fourier transform (FT). The velocity curve is created from the weighted average of the Doppler shifts, found by individually fitting the H$\beta$-H$\epsilon$ lines with a combination of a Gaussian and Lorentzian function fixed to have the same central wavelength. The average noise level of the velocity FT is 0.56~\kms. See Figure~2.

We fit and remove individual sine waves to the light curve starting with the largest amplitude mode, the order is depicted in Table~1. We then fit modes that represent combinations of these larger modes. Overall, we find eight independent frequencies and five combination frequencies in the light curve. The final fit in Table~1 represents fitting all 13 modes simultaneously. The errors represent analytical estimates of the uncertainties calculated by Period04 \citep{p04,breger99}, a program that performs FTs and sine-wave fits to a given light curve. The reduced chi-square value for the fit to the light curve is 2.4, indicating that our 13 modes do not represent the entire signal in the light curve given the size of the error in each light curve measurement. The remaining signal is dominated by low-frequency signals that are typical of variable seeing and are probably not inherent to the star.  The nonlinear light curve of G~29-38 also suggests that combination modes below the noise level are present proving further signal that we did not remove. The errors presented in Table 1 represent this reduced chi-squared value.

The strongest periodicity of the velocity curve corresponds to the strongest mode (F1) found in the light curve, giving us confidence that we have measured the motion of the gas on the surface. The light curve measures the period of each mode better than the velocity curve because of its lower noise level. We fit the velocity curve with sine waves, fixing the frequencies to the thirteen modes measured in the light curve. The reduced chi-square for the fit of the 13 frequencies to the velocity curve is 1.6. The remaining signal in the velocity FT is at low frequencies and probably result from stellar wander in the wide slit. The lower reduced chi-square of the velocity fit as compared to the light curve fit indicates that the velocity curve is more linear, i.e. no undetected combination frequencies. The amplitudes and frequencies found by fitting the velocity and light curves at the same frequencies are marked on the FTs in Figure 2.

To determine the significance of a mode we use the false alarm probability \citep{kepler93}. In the light curve, to have less than 0.1~\percent\ chance of a peak being produced by noise alone it must have an amplitude of at least 3.2~mma.  For the combination modes, where we know the frequency lies at the combination of the two dominant modes, the peak need only have an amplitude of 2.3~mma.
Table~1 only includes peaks in the FT with amplitudes greater than these values. For the velocity FT, peaks need an amplitude of 2.04~\kms\ to be detected independent of the light curve with the same 0.1~\percent\ confidence level. However, since the light curve fixes the frequency, for a velocity mode to have less than 0.1~\percent\ chance of being due to noise it simply needs an amplitude greater than 1.47~\kms. Only the modes F1, F2 and the combination of F1+F2 have velocity amplitudes greater than this limit. 

We compare the observed velocity and light amplitudes and phases with the values $R_v=$(A$_v$/$\omega$)/A$_f$, relative amplitude, and $\Delta\Phi=\Phi_v-\Phi_f$, relative phase in Table~\ref{t:G29}. The measured $R_v$ values have been shown by \citet{thompson03} and VK00 to generally agree with the convective driving theories of \citet{gw1}.  Our measurements of G~29-38 agree with the previous measurements with the Keck telescope. One mode present in all three data sets (615~s) consistently has the same relative amplitude: R$_v$ is 17$\pm$3, 12$\pm$4, and 18$\pm$2 Mm/rad for VK00, \citet{thompson03} and this paper respectively. These measurements are all near to the prediction by \citet{gw2} of 16~Mm/rad for an 800~s mode on a similar temperature star.  The relative phases, $\Delta\Phi$, for those modes with a significant velocity detection all show the expected lag of the brightness behind the motion of the gas, predicted by \citet{gw2} to be a 55$^\circ$ lag for an 800~s mode.

\begin{figure}
\includegraphics[scale=.65, angle=270]{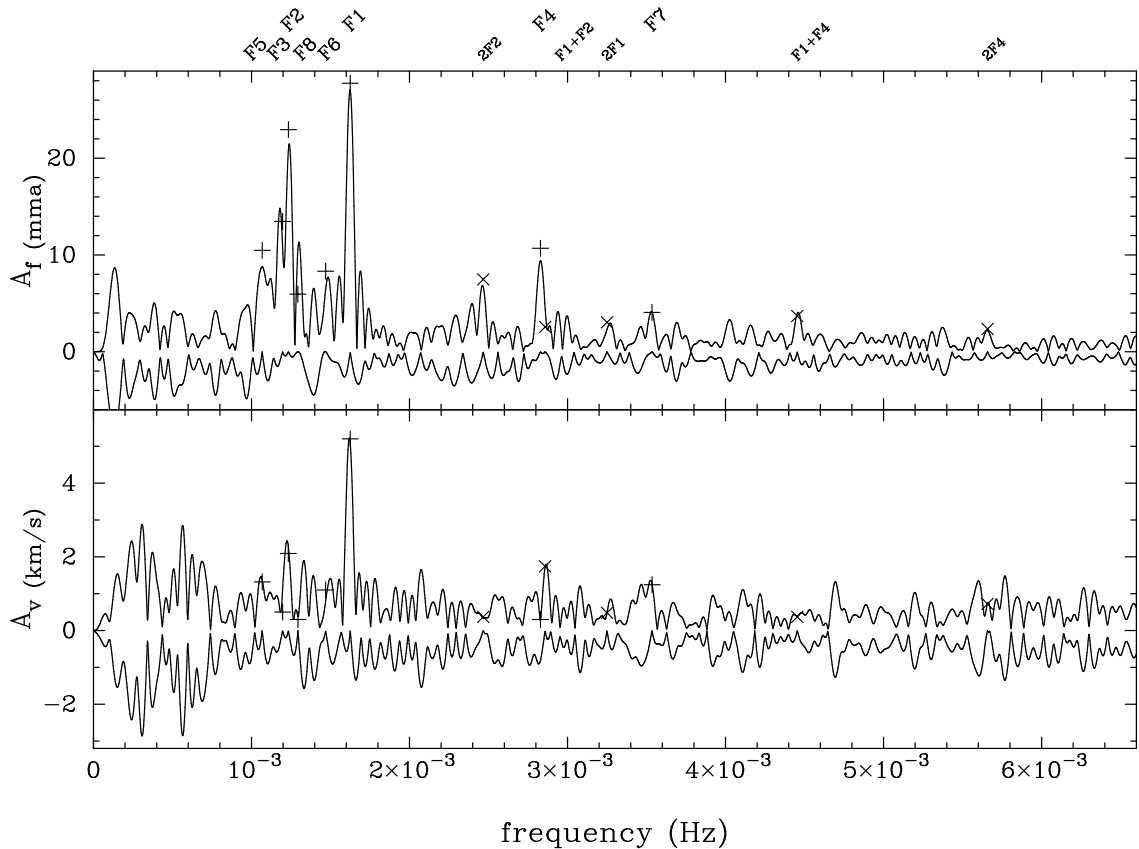}
\caption{The Fourier transforms of the light (top) and velocity (bottom) curves
of G~29-38. The modes found in the data are labelled at the top.
Sine-wave fits to each curve at these frequencies are marked on the FT. Pluses represent independent modes and crosses represent combination frequencies. The residuals after the fit are shown as reflections beneath each FT.}
\end{figure}

\begin{table*}
%\typesize{\footnotesize}
\begin{minipage}{500pt}
\caption{The modes measured in the light and velocity curves of G~29-38. The fits to the flux time series use the function
$A_f cos(2\pi ft-\Phi_f)$ and the velocity time series fits use the function $A_v cos(2\pi ft-\Phi_v)$. Comparison of the 
flux to the velocity pulsations are $R_v=(A_v/\omega)/A_f$ and $\Delta\Phi=\Phi_v-\Phi_f$. The comparison of the combination modes to their parent modes are given as $R_c=A_f^{i+j} / n_{ij}A_f^iA_f^j$ and $\Delta\Phi_c=\Phi_f^{i+j}-(\Phi_f^i+\Phi_f^j)$. \label{t:G29}}

\begin{tabular}{lcccccccccc}
 & $P$ & $f$ & $A_f$ & $\Phi_f$ & $A_v$ & $\Phi_v$ &$R_v$ &$\Delta\Phi$ &R$_c$ & $\Delta\Phi_c$\\
 &\small(s) & \small($\mu$Hz) & \small(mma) & \small($^\circ$) & \small(\kms) & \small($^\circ$)
 & \small($\frac{Mm}{\%}$) & \small($^\circ$) &  & \small($^\circ$) \\

\hline
\hline
%\startdata
F1 &615 & $1625.0 \pm0.9$  &$27.7 \pm1.0 $&$226 \pm2  $&$5.2 \pm0.5 $&$286 \pm5  $&$18.3 \pm1.7 $&$ 60 \pm5$ & -- & --\\
F2 &811 & $1233.4 \pm1.0$  &$22.9 \pm1.0 $&$189 \pm2  $&$2.1 \pm0.5 $&$225 \pm13 $&$11.8 \pm2.8 $&$ 36 \pm13$ & -- & --\\
F3 &835 & $1197.2 \pm1.0$  &$13.4 \pm1.0 $&$192 \pm4  $&$0.5 \pm0.5 $&$102 \pm55 $&$4.9  \pm4.8 $& --  & -- & --\\
F4 &353 & $2829.1 \pm2.1$  &$10.7 \pm1.0 $&$82  \pm5  $&$0.3 \pm0.5 $&$52  \pm54 $&$1.6  \pm2.5 $& --  & -- & --\\
F5 &937 & $1067.6 \pm2.1$  &$10.5 \pm1.0 $&$290 \pm5  $&$1.3 \pm0.5 $&$33  \pm20 $&$18.8 \pm6.8 $&$ 103\pm21$ & -- & --\\
F6 &681 & $1469.3 \pm3.0$  &$8.3  \pm1.0 $&$274 \pm6  $&$1.1 \pm0.5 $&$341 \pm24 $&$14.3 \pm6.2 $&$ 67 \pm25$ & -- & --\\
F7 &283 & $3534.5 \pm5.8$  &$4.1  \pm1.0 $&$347 \pm13 $&$1.2 \pm0.5 $&$37  \pm20 $&$13.8 \pm6.1 $&$ 51 \pm24$ & -- & --\\
F8 &773 & $1293.9 \pm4.3$  &$5.9  \pm1.0 $&$238 \pm10 $&$0.3 \pm0.5 $&$23  \pm90 $&$6.3  \pm9.6 $& --  & -- & --\\

2F1  &308 &  $3250.1 \pm0.0$  &$3.0 \pm1.0 $&$76 \pm19 $&$0.48 \pm0.5 $&$180 \pm55  $&$7.7 \pm7.8  $& -- & $3.9\pm1.3$ & $-16\pm19$\\
2F2  &405 &  $2466.8 \pm0.0$  &$7.5 \pm1.0 $&$356  \pm7  $&$0.37 \pm0.5 $&$83  \pm70  $&$3.2 \pm4.0 $& -- & $14.1\pm2.1$ & $-22\pm8$\\
2F4  &177 &  $5658.2 \pm0.0 $  &$2.3 \pm1.0 $&$110 \pm23 $&$0.72 \pm0.5 $&$347 \pm36  $&$8.6 \pm6.5  $& -- & $20.6\pm9.2$ & $-54\pm 24$\\
F1+F4 &224 & $4454.1 \pm0.0$  &$3.7 \pm1.0 $&$283 \pm15 $&$0.37 \pm0.5 $&$114 \pm71  $&$3.6 \pm4.5  $& -- & $6.2\pm1.8$ & $-25\pm 16$\\
F1+F2 &350 & $2858.4 \pm0.0$  &$2.6 \pm1.0 $&$70 \pm24 $&$1.74 \pm0.5 $&$87 \pm15  $&$37.8\pm17.2  $&$17 \pm28$ & $2.0\pm0.8$ & $15 \pm 24$\\
%\hline
%\enddata
\end{tabular}
\end{minipage}
\end{table*}

With the observation of several combination modes, we compare the combination amplitudes and phases with their parent modes according to VK00 and \citet{wu01}. Specifically, we show the combination amplitude ratio, $R_c=A_f^{i+j} / n_{ij}A_f^iA_f^j$, and the combination phase difference, $\Delta\Phi_c=\Phi_f^{i+j}-(\Phi_f^i+\Phi_f^j)$, in Table~1. The $R_c$ values for the first harmonics of our F1 and F2 agree within the errors with the same modes presented by VK00 on G~29-38.  Since combination modes do not suffer from cancellation over the surface of the star in the same manner as their parent modes, the value of $R_c$ can be an indication of $\ell$ \citep{yeates05, wu01}. However, these values are dependent on the azimuthal order, $m$, and the inclination of the star. Since we have not resolved rotational splitting of our modes, use of combination modes to determine $\ell$ is very uncertain.  We only note that the first harmonic of F4 shows a larger value than the other harmonics.    

One curious feature in these time series is the modestly significant peak in the velocity FT at 2858$\mu$Hz that matches the combination frequency F1+F2. While the combination of F1+F2 lies within 30$\mu$Hz of F4, it is impossible to satisfactorily fit the observed velocity peak with the same frequency as F4. While the 1.74\kms amplitude of this velocity curve is above the significance level, this criterion only applies if we can expect to find combination frequencies in the velocity curve. Otherwise the false alarm probability is 1/26, i.e. a 3.8~\percent chance of the velocity peak being due to noise. 

Most observations and theory insist that combination modes should have no motion corresponding to the flux variations \citep{wu01,vk00}.  Combinations are believed to occur from the changing depth of the outer convection zone, attenuating, delaying and distorting the pulsations \citep{montgomery05, wu01}.  Hence we observe nonlinear pulse shapes in the light curve and combination modes in the brightness FT.  Since the combinations occur from nonlinear mixing of the temperature variations, they do not require gas motions to be seen as a peak in the light FT.  However, \citet{thompson03} previously found one instance of a significant velocity variation on G~29-38 that matches the sum of two large modes. So, despite the theoretical predictions, observations tell us that we can expect to find velocity combination and we conclude that F1+F2 is the second observed combination mode demonstrating an associated periodic velocity variation.

\section{Mode identification of G~29-38}
Due to differences in limb darkening across the spectral lines, the hydrogen spectral line shapes change according to the distribution of the pulsation across the surface of the star. As a result, chromatic amplitudes, wavelength dependent amplitude variations across the Balmer lines, have characteristic shapes that reveal $\ell$.  This technique cannot reveal $m$ because of the purely spherical harmonic treatment of the brightness variations \citep{robinson95}. Also, the motion of the gas, which is variable over the surface of the star and dependent on $m$, is too small to observe except for the average change of the line center as shown in the previous section.

\begin{figure}
\includegraphics[scale=.65, angle=270]{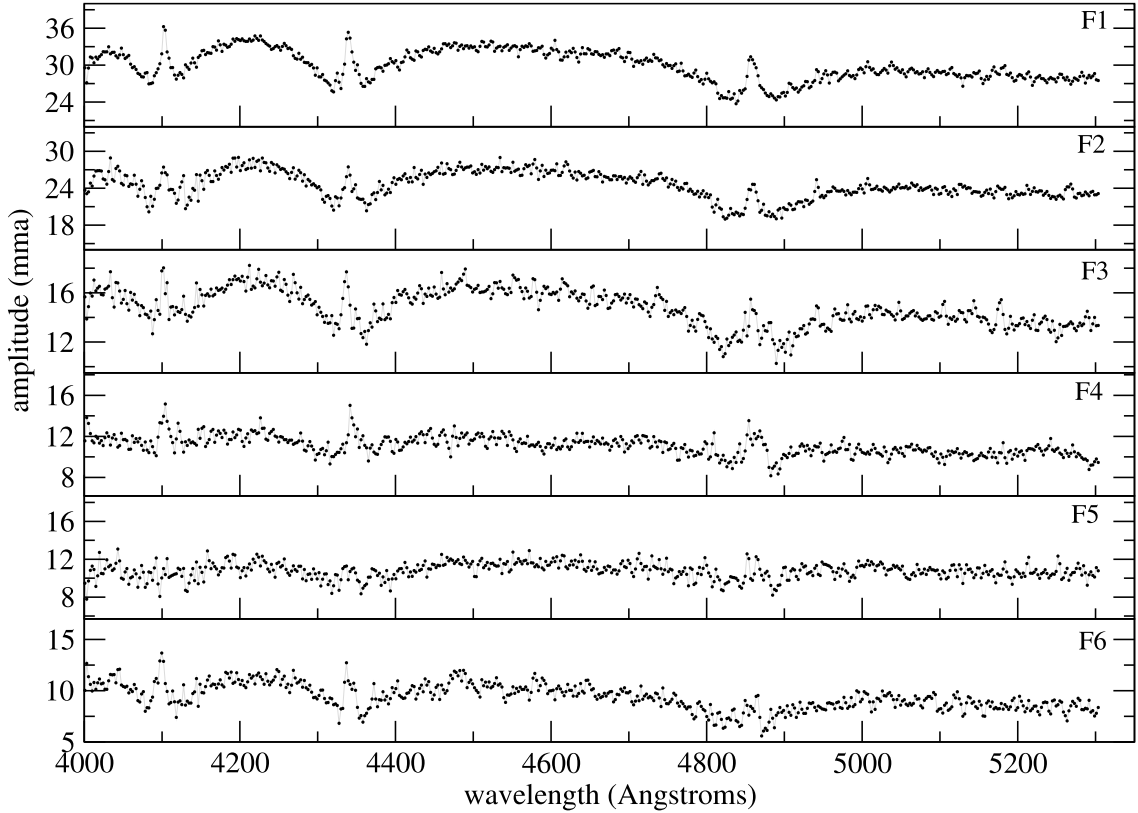}
\caption{The chromatic amplitudes of the six largest pulsation modes
present in the VLT data. These chromatic amplitudes are created from Gaussian smoothed ($\sigma=2.5$\AA) spectra. The smoothing improves the signal to noise in each wavelength
bin.\label{f:rawca}}

\end{figure}

Using the technique of chromatic amplitudes, C00 show that one mode has $\ell$=2 while all others match $\ell$=1 on G~29-38. We present the chromatic amplitudes from the VLT data of the six largest pulsation modes in Figure~\ref{f:rawca}.  As seen across the H$\beta$ through H$\delta$ lines, our data demonstrate the typical central peaks with side dips predicted by the atmospheric models for low values of $\ell$.  However, the noise, especially in the lower amplitude modes, prevent us from making specific mode identifications when we compare them to the models. One obvious feature is that the 353~s mode, F4, appears to have shallower dips on either side of its peaks, indicating that F4 is a different spherical degree than F1 or F2. F5 also seems to have somewhat shallower dips, though the increased noise makes the difference much less significant.

Since the introduction of chromatic amplitudes, fitting spectral lines has proven to be useful in extracting spherical degree from lower quality data \citep{thompson04}. As a result, we fit the spectral lines of our VLT data to measure the spherical degree of the largest amplitude pulsations. We apply the same technique to the previous Keck observation of G~29-38 (VK00, C00) to prove that our method produces the same results as the raw chromatic amplitudes and to be able to directly compare the measurements of the two data sets.

\subsection{How to measure spherical degree}
\label{s:ew}
We follow the procedure outlined by \citet{thompson04} for fitting each spectral line with the combination of a Gaussian and a Lorentzian superimposed on a linear continuum. We fit each spectrum only allowing the area of the Gaussian and Lorentzian to vary along with the slope and intercept of the linear continuum. Prior to fitting we establish the initial conditions of all the line parameters by fitting the average spectrum; this established the full-width-half-max values of both functions and the central wavelength of the line.

As previously discussed, one method to present the line shape variations is to create chromatic amplitudes by measuring the amplitude of each pulsation mode for each wavelength bin. In this case we use the fitted spectra to create the chromatic amplitudes. Figure~4 shows the chromatic amplitudes for a temperature and gravity model consistent with G~29-38. The fits to the model spectral lines are over a limited wavelength range and do not measure the amplitudes between the spectral lines. As expected each spherical degree has a different chromatic amplitude shape.

Alternatively, since we fit the spectra, we can reduce the problem of distinguishing $\ell$ to quantities measured from the varying fitted parameters. Specifically, we measure the fractional amplitudes of the fitted EWs for the Gaussian (G$_{EW}$) and the Lorentzian (L$_{EW}$) functions at the known periods of pulsation. 
We also measure the fractional light amplitude for each pulsation mode from the flux measured by the linear continuum of the fit at the central wavelength of the line (A$_{\lambda}$). These flux amplitudes agree within the errors with those measured directly from the spectra in Table~1.  We normalize the EW amplitudes by this light amplitude for each mode and plot G$_{EW}$/A$_\lambda$ against L$_{EW}$/A$_{\lambda}$ for each spectral line in Figure~4.

For the data and models considered here we fit across 200~\AA\ for the H$\beta$ and H$\gamma$ lines and 100~\AA\ for the H$\delta$ line. All Doppler shifts of the spectral line have been removed prior to fitting the spectra to reduce the number of parameters in the fit, leaving the four previously mentioned fitted parameters. 
Generally, the Lorentzian fits the wide wings of the spectral lines while the Gaussian is limited to the central core.  The Gaussian represents only about 3~\percent of the line area, however it also shows larger fractional variations than the Lorentzian.
 
Using the limb-darkened models of DA white dwarf stars \citep{weko84}, we pulsate the models with a spherical harmonic according to $\ell$=1,2,3 and 4.   Each model is Gaussian smoothed by $\sigma=2.9$~\AA\ to account for the seeing that reduces the resolution of the spectra, though this has only a small effect on the results. We force the fits of the model spectra and the observations to have the same ratio between the full-width-half-max of the Gaussian and the Lorentzian functions. In this way we further control the fit, providing consistent results between the observations and the model. We choose a model of T=11,750~K and log(g)=8.25 (cgs), appropriate given the temperature and gravity measured by VK00.

Fits to model spectra show that different values of $\ell$ will produce unique Gaussian and Lorentzian EW amplitudes. A comparison of the chromatic amplitudes and the EW variations in Figure~4 show the correlation between the two methods of presenting the periodic line shape variations. The amplitude of the Gaussian EW corresponds to the central bump seen in the chromatic amplitudes, while the Lorentzian EW corresponds to the wings of the chromatic amplitudes. With only two parameters, the EW variation plot simplifies the comparison between observed pulsations and the models.  In this paper we use the normalized EW variation plots to quantify the Keck, VLT and model line shape variations.

\begin{figure}
\includegraphics[scale=.33]{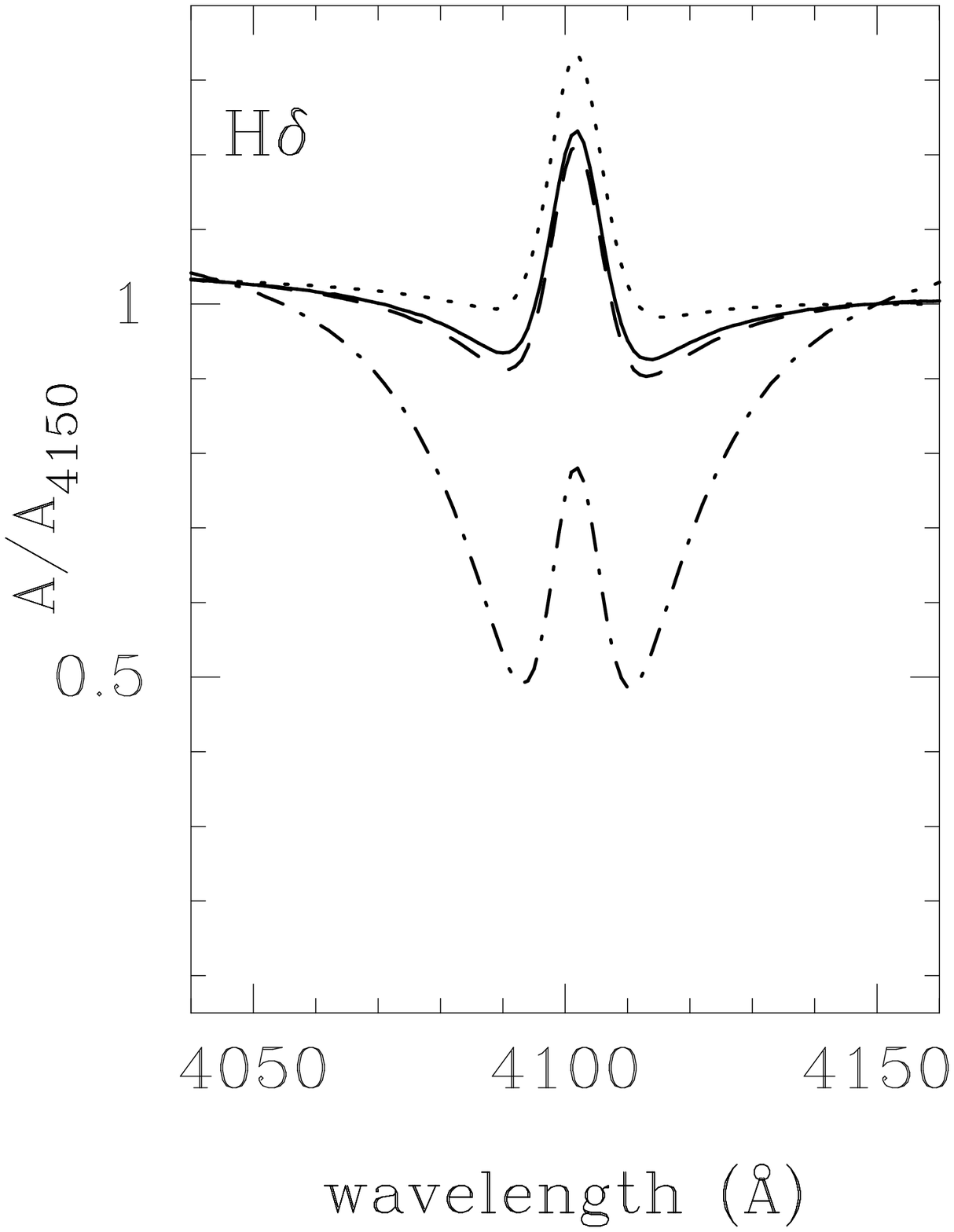} 
\includegraphics[scale=.33]{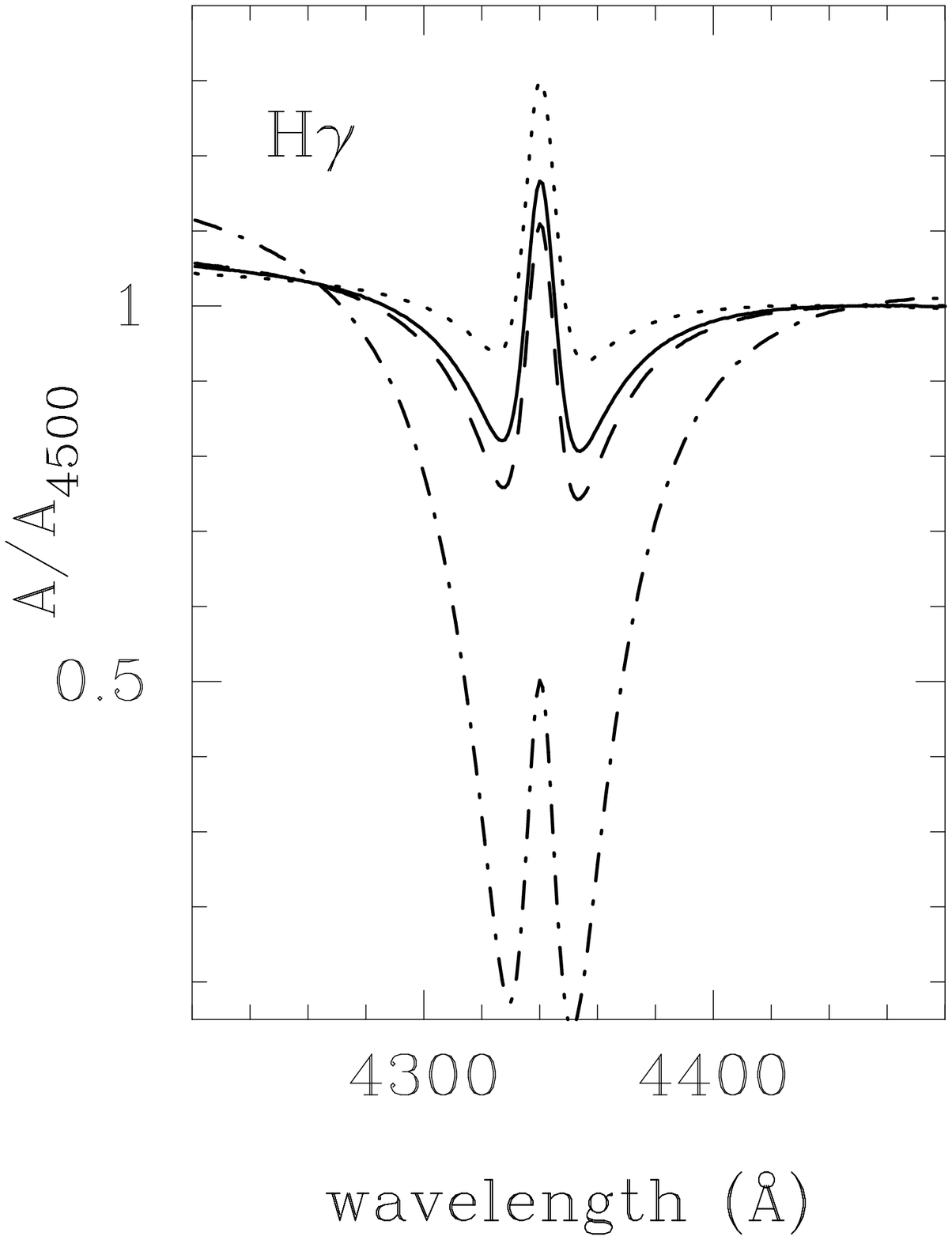}
\includegraphics[scale=.33]{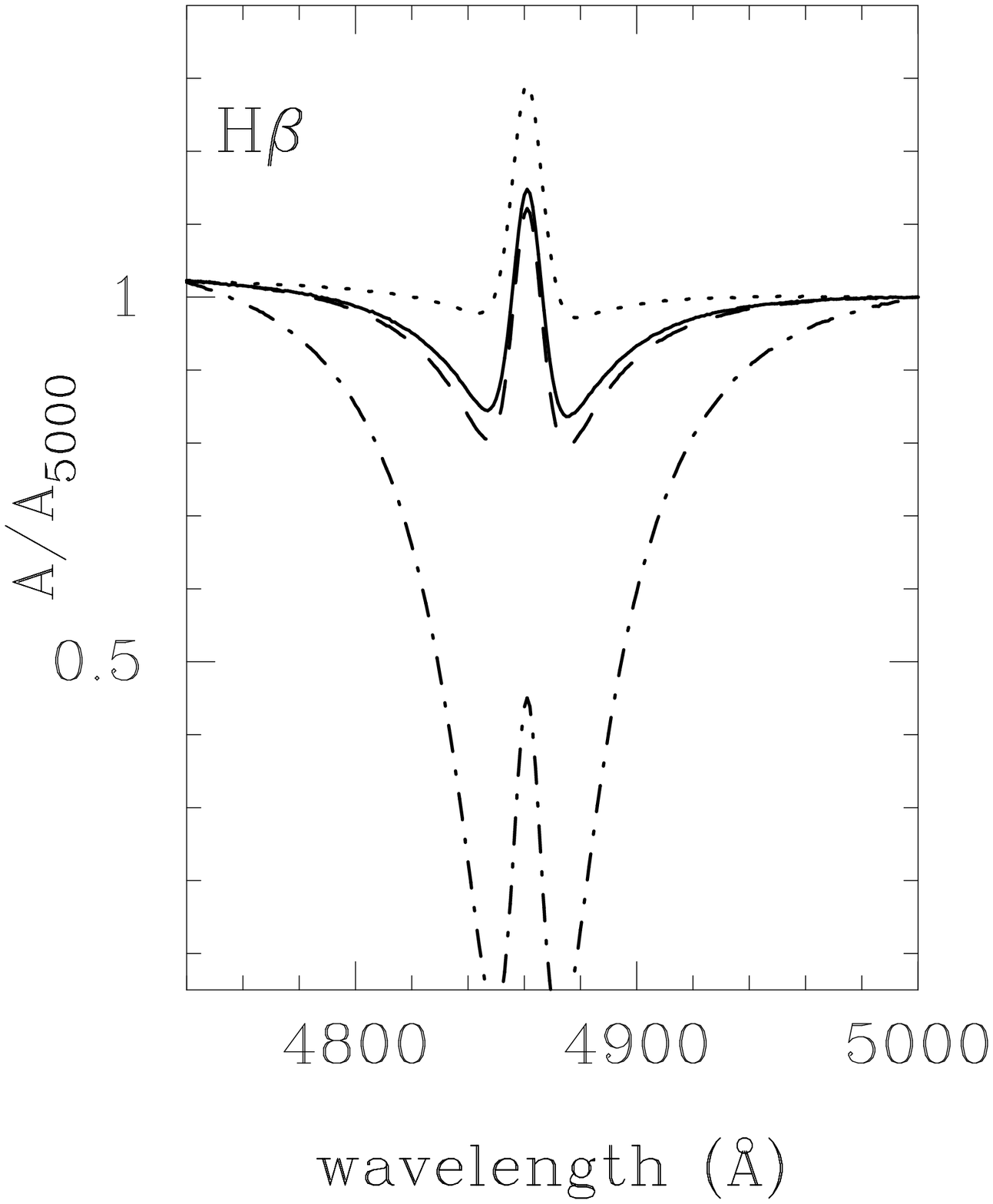}

\

\includegraphics[scale=.33]{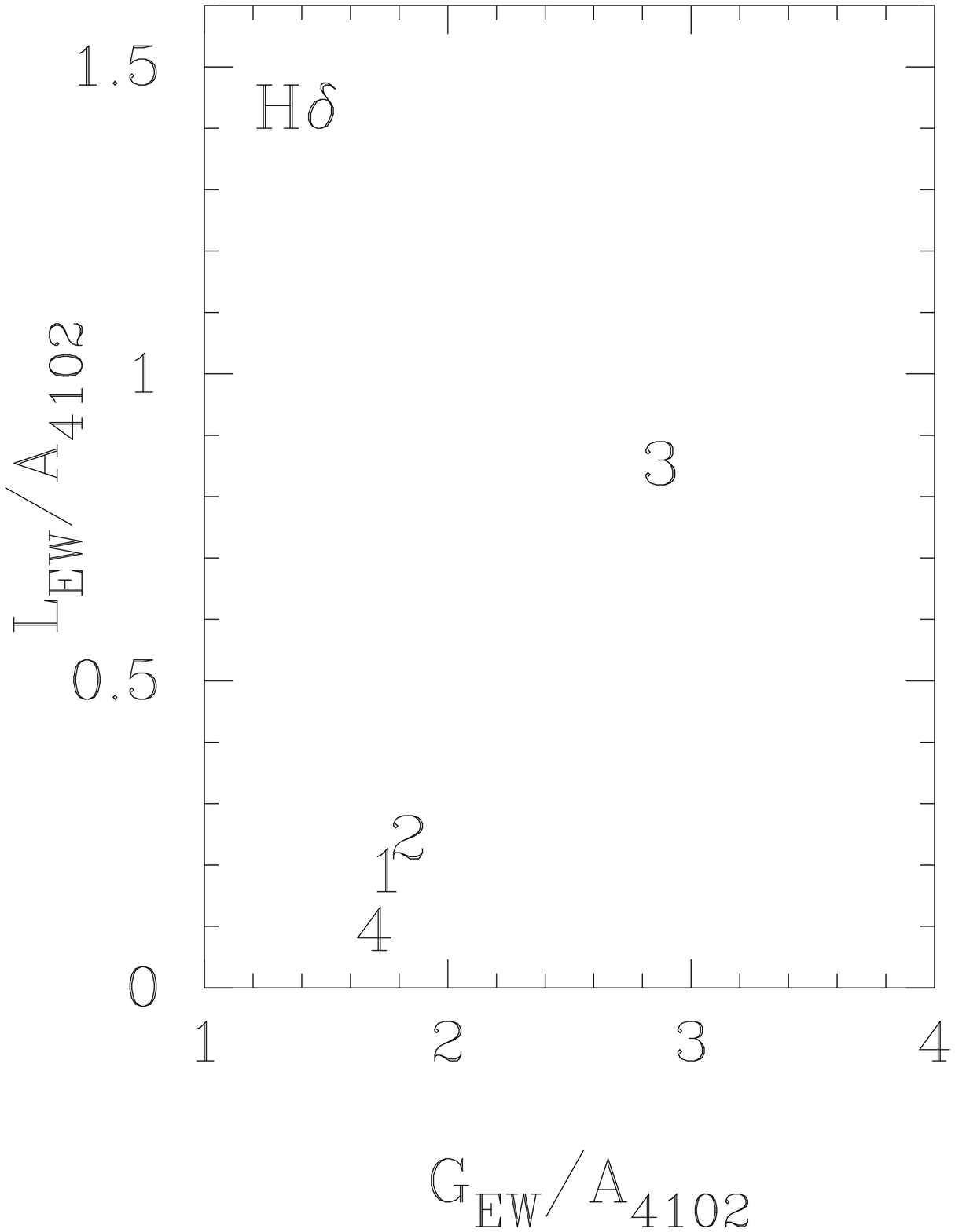} 
\includegraphics[scale=.33]{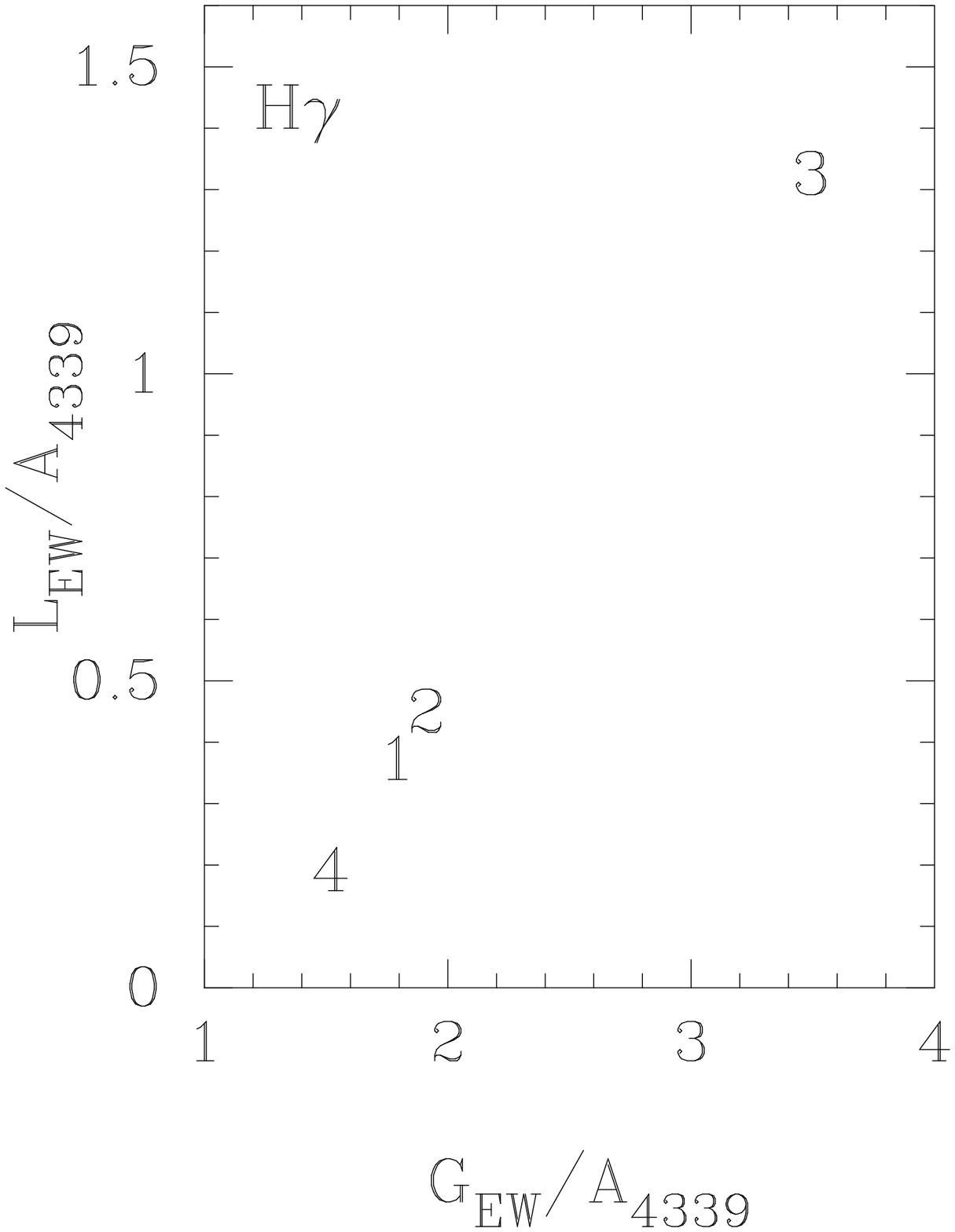}
\includegraphics[scale=.33]{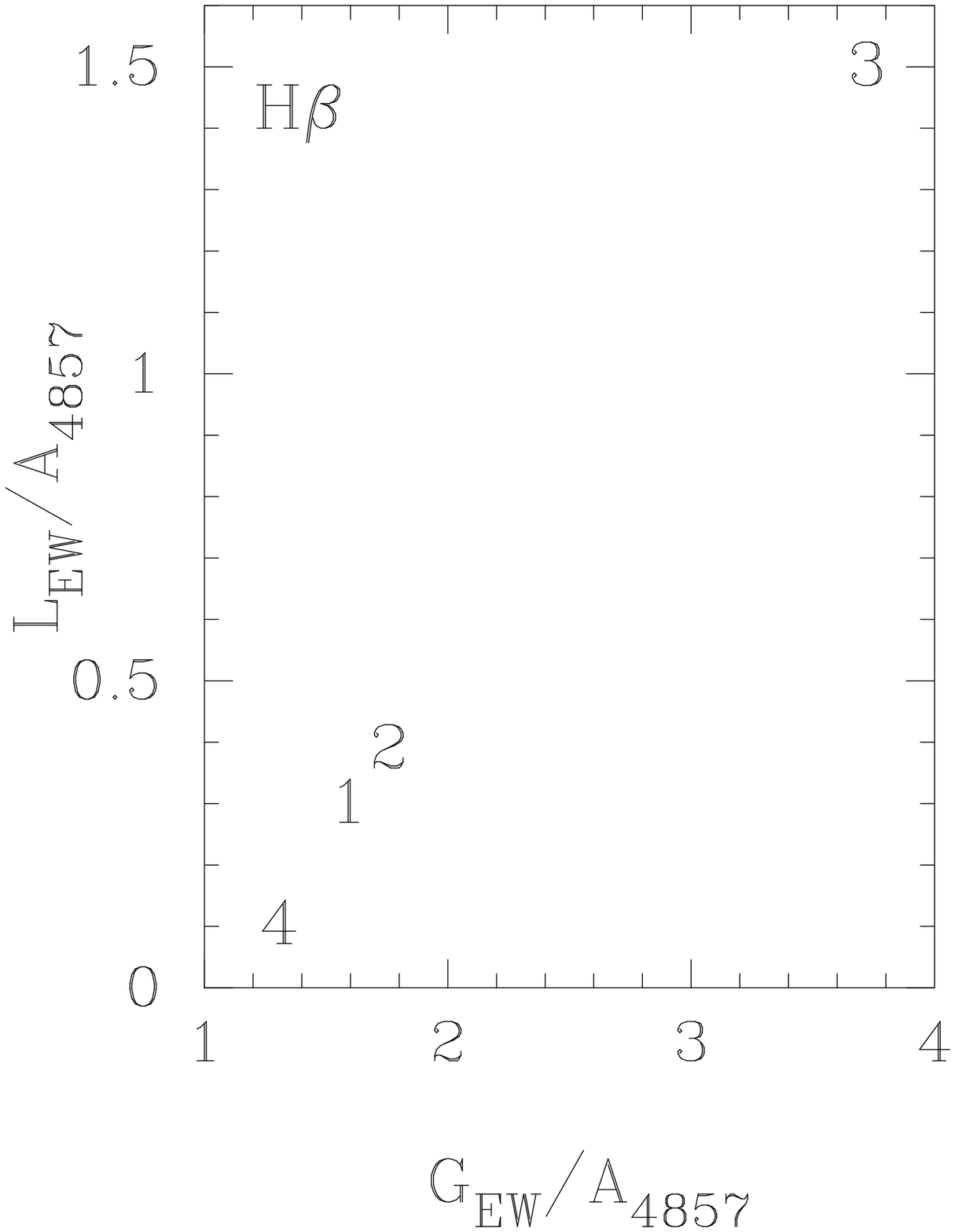}
\caption{Analysis of model time series spectra for T=11,750K and log(g)=8.0 pulsated with $\ell$=1,2,3 and 4. Each frame represents the fit to a different hydrogen Balmer line. Top figures are chromatic amplitudes. Bottom are Gaussian vs. Lorentzian EW fractional amplitudes (G$_{EW}$ and L$_{EW}$) normalized by the light fractional amplitude (A$_\lambda$) measured for each line. See \S\ref{s:ew} for details.}
\label{f:model}
\end{figure}

\subsection{Measured EW variations}

We measure the periodic changes in the normalized EWs for the six largest modes of the VLT observations and perform the same analysis on the Keck time series spectroscopy of G~29-38 (VK00). With a consistent analysis we can compare the data sets to the models as well as to each other.  Figures~5 and~6 show analysis of both data sets; they are consistent but show some offset from the model. The presented error bars are the formal errors as measured from sine wave fits at the stated period to the time series of the EWs and continuum fluxes (in the next section we describe monte carlo simulations that demonstrate these errors are reasonable.) 

Using chromatic amplitudes measured straight from the data C00 discovered five $\ell$=1 modes and one $\ell$=2 mode. Instead, we fit the spectra to create EW variation plots for the 6 largest, real modes they measured in their light curve. Our results from fitting the spectra agree with the previous conclusions of C00. Figure~5 shows these results along with the model. While a small offset between the model and the Keck observations exist, the general trends between different spherical degree appear to be accurate. These trends are independent of the model temperature and gravity and thus, we can still determine $\ell$ because we can compare observed modes. Specifically, the mode labelled `d' is the mode that C00 measured as $\ell$=2 (776~s). It has larger variations in the Lorentzian and Gaussian EW than the other observed modes, exactly as seen in the model. 

Our results from fits to the VLT spectral series taken of G~29-38 
are shown in black in Figure~6. Only the six modes with the largest light amplitudes are shown. The majority assuredly overlap with the cluster of modes observed with the Keck data (circles). Specifically, our F1 and F2 are the same period as `a' and `b' in Figure~5, and they agree in value within the error bars. The majority of modes in G~29-38 are expected to be $\ell$=1 \citep{kleinman98}, and our observed cluster in the Keck and VLT observations agree with this conclusion.

We note two interesting modes in the VLT data. One is the 681~s mode, labelled `F6'. It consistently shows larger EW amplitudes in all the spectral lines, just like the previously known $\ell$=2 mode (`d' in Figure 5). The other feature is the 353~s mode, labelled `F4'. It consistently shows smaller EW variations, similar to what might be expected of an $\ell$=4 mode. We note that this mode lies very near to the combination of the two primary modes (F1+F2) and the small combination could be influencing our results. However, simulations that include nearby combinations modes have no effect on the measured EW variations.

\begin {figure}
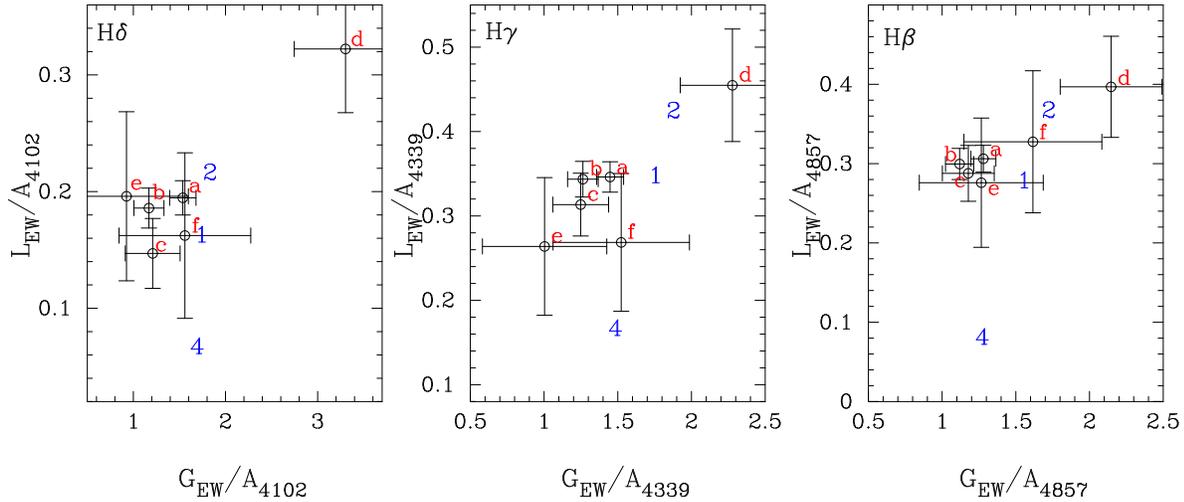

\includegraphics[scale=.33]{figure5a.ps} 
\includegraphics[scale=.33]{figure5b.ps} 
\includegraphics[scale=.33]{figure5c.ps}
\caption{Normalized EW amplitudes of each spectral line (H$\beta$, H$\gamma$ and H$\delta$) of the Keck data as compared to the model. Similar to Figure~\ref{f:model}. The model is labelled according to the spherical degree it represents (1, 2 and 4).  The Keck modes are:  a=614~s, b=818~s, c=653~s, d=776~s, e=283~s, f=430~s. The mode
labelled 'd' is the one claimed by C00 to have a spherical degree of two.} 
\end{figure}

\begin {figure}
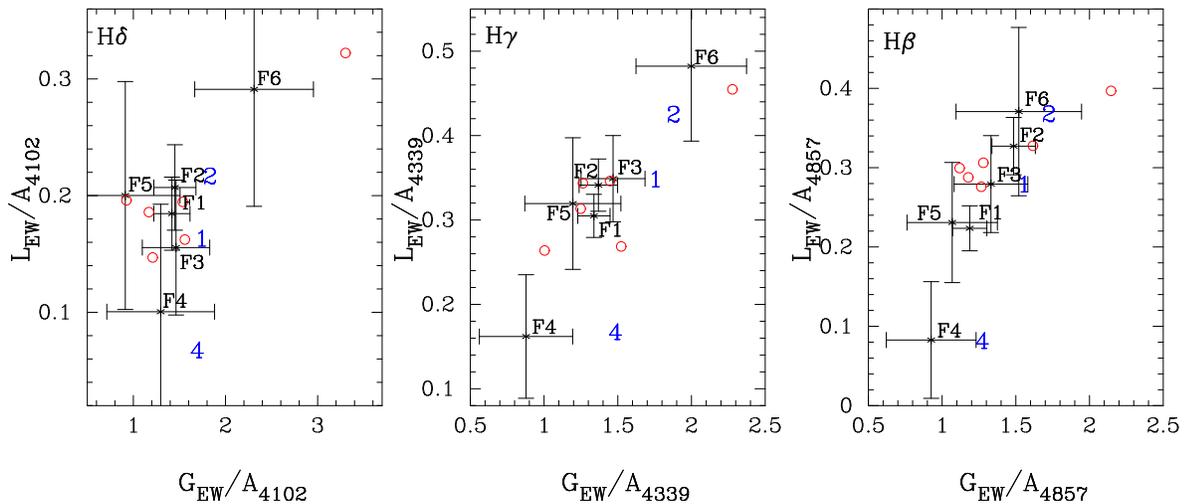

\includegraphics[scale=.33]{figure6a.ps} 
\includegraphics[scale=.33]{figure6b.ps}
\includegraphics[scale=.33]{figure6c.ps}
\caption{Normalized EW amplitudes of the H$\beta$, H$\gamma$ and H$\delta$ spectral lines for the VLT observations (black error bars) as compared to the model (blue numbers) and the Keck measurements (red circles). Similar to Figure~\ref{f:model}.
See Table~1 for a list of periods associated with each labelled mode.}
\label{f:mcarlo}
\end{figure}

\subsection{Noise simulations}
To better understand how noise affects our measurement of spherical degree, we perform Monte Carlo simulations of model spectral series. Gaussian noise was introduced to both the overall brightness of the spectra and across the individual wavelengths of the model spectrum to match the noise observed in the VLT data. The error produced in individual fits is equivalent to the error  observed in the same amplitude mode in the VLT data, confirming that we are closely emulating the observations. We produce and fit the H$\gamma$ line of 500 simulated spectral series (with 604 spectra in each series) and plot regions containing the central 68~\percent\ of the points for each $\ell$.  We show this scatter in Figure~\ref{f:mcarlo} for amplitudes of 27~mma, 10~mma and 8~mma, consistent with modes F1, F4 and F6 respectively. 

From the simulations for the VLT noise level, we expect to be able to distinguish $\ell=1$, $2$ and $4$ for the largest amplitude modes even if we only measured the gamma line.  Given the simulated scatter, we would not be able to distinguish F6 from a single spectral line. However, the fact that we see the same trend in all three lines, increases the probability that we correctly identified all our modes.  The simulations show that the normalized Gaussian and Lorentzian EW amplitudes are correlated,  but the size of the error bars are basically reflected by the 1$\sigma$ error bars presented in Figures 5 and 6.  

\begin{figure}
\includegraphics[scale=0.34]{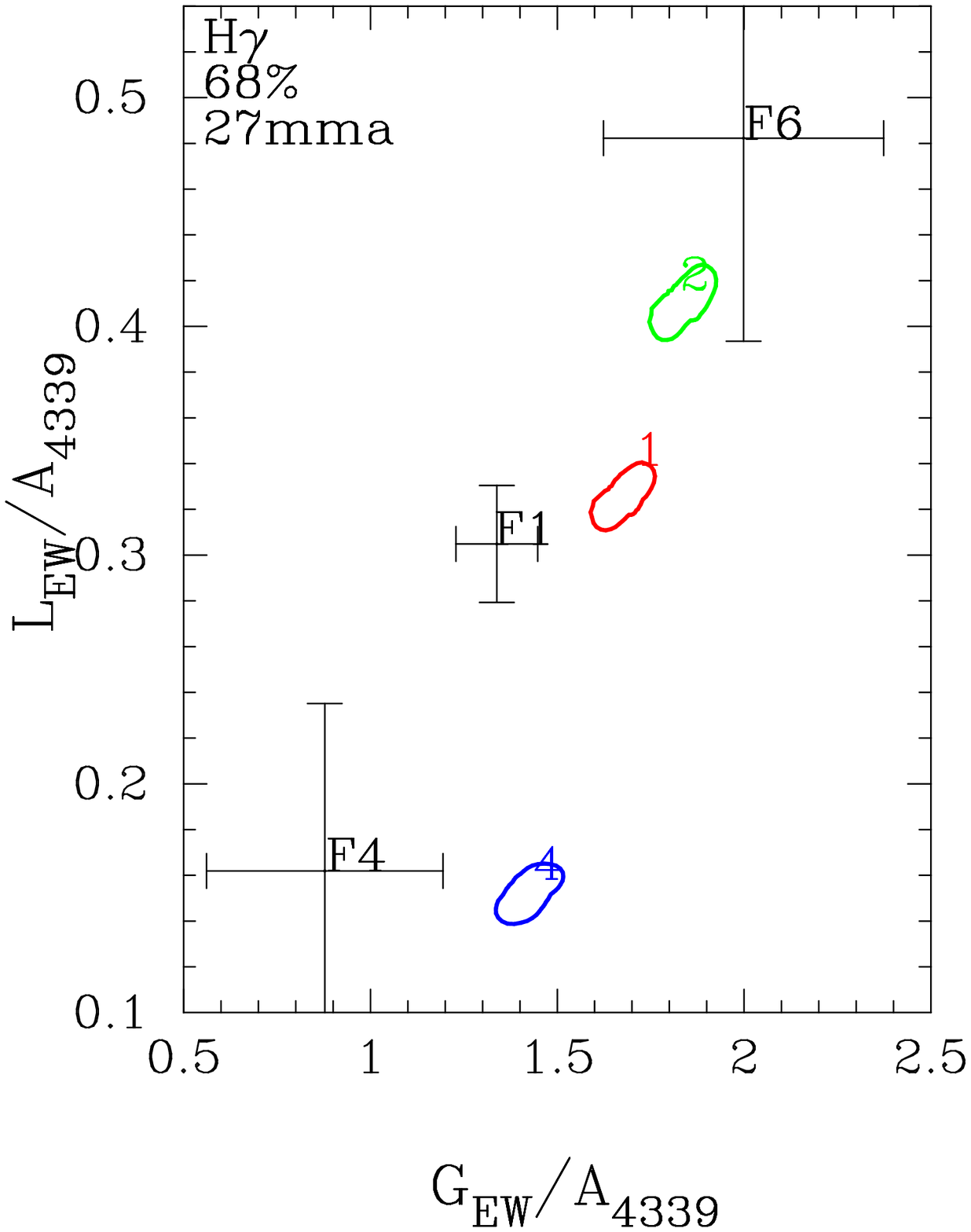}
\includegraphics[scale=0.34]{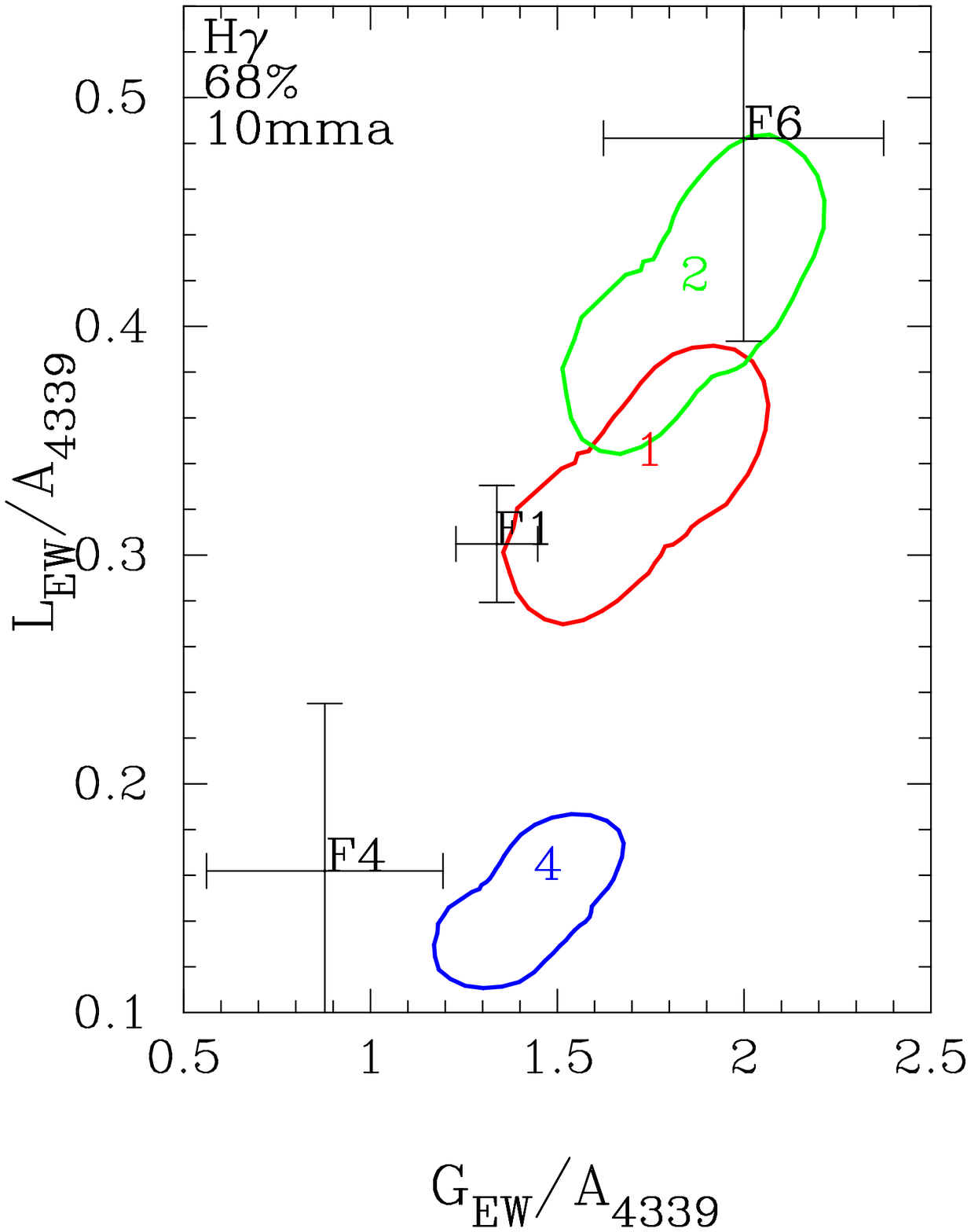}
\includegraphics[scale=0.34]{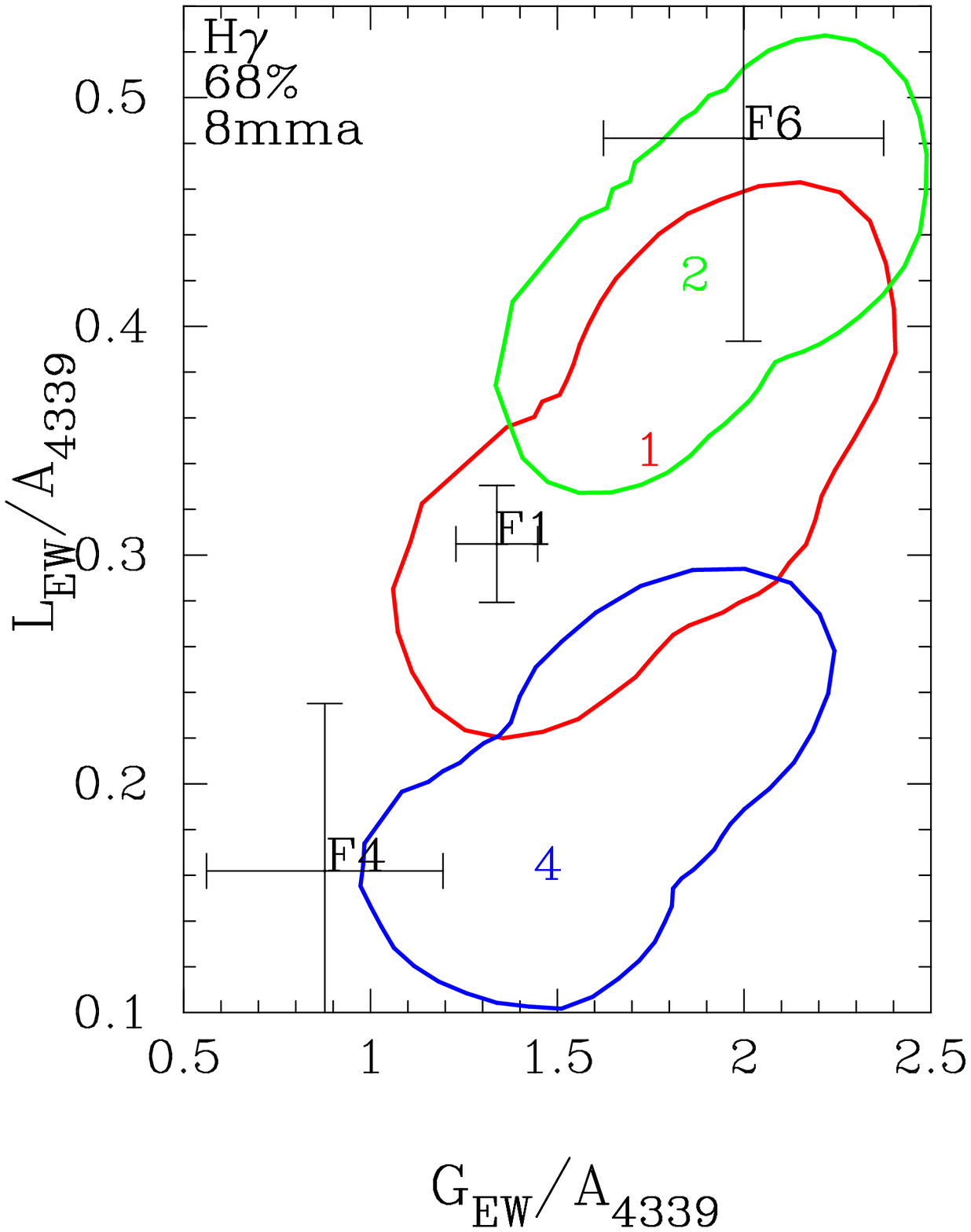}
\caption{Results of a 500 iteration monte carlo simulation of fits to H$\gamma$ of model spectra emulating the noise of the the VLT observations.  For each $\ell$, 68~\percent\ of the 500 iterations lie in the area outlined. The noise free model is
labelled by the number of its $\ell$\ value. F1, F4 and F6 of the 
VLT measurements are included for comparison.  
The extent of the scatter is similar to the error bars of the equivalent amplitude mode, however the Gaussian and Lorentzian measurements are correlated.}  
\end{figure}

\subsection{Discussion}
\label{s:ewd}
Due to the limb-darkening variation across the spectral lines, periodic changes in the line shape reveal the distribution of the pulsation across the surface of the star. The observations of the line shape do not explicitly agree with the model, as seen in previous chromatic amplitudes (C00). However, the trends between different spherical degrees do agree and as a result, we have measured $\ell$ on G~29-38. The F1, F2, F3 and F5 are all $\ell=1$. F6 (681~s), with its larger EW amplitudes, is consistent with the $\ell$=2 model and the previously identified $\ell$=2 mode at 776~s (C00). F4 shows decreased EW amplitudes in each spectral line, indicative of $\ell$=4.

We note two reasons to be cautious of F6's identification as $\ell$=2. First, F6 does not show the large R$_v$ value seen by VK00 for their $\ell$=2. Considering the error on the flux and velocity amplitudes, this could simply be the result of noise. Second, the 2$\sigma$ error bar of F6, as determined from the monte carlo simulations, includes the $\ell$=1 model. The observed increase in EW amplitudes of one spectral line could simply be due to the scatter. However, since we see the same trend in all three spectral lines, it is likely that this is a real effect and we have correctly identified F6.

F4 (353~s) consistently has smaller EW variations in each spectral line than every other mode, similar to the models of $\ell$=4.  Only one other $\ell$=4 mode has been measured on a DAV.  In that case the spectroscopic evidence was bolstered by UV observations and combination amplitudes \citep{thompson04}. Like our observations, this $\ell$=4 on G~185-32 shows reduced Gaussian and Lorentzian EW amplitudes as compared to the other modes \citep{thompson06}. The similarity between our observations of F4 and this established $\ell$=4 mode gives strength to the claim that F4 is $\ell$=4.

Another indication of higher $\ell$ is the presence of large amplitude harmonics as compared to the parent modes. \citet{yeates05} use the theories of \citet{wu01} along with the amplitudes of combination modes to show that G~185-32 does indeed have a higher $\ell$ mode, though their results better agree with $\ell$=3. Here, F4 has harmonic modes despite its low amplitude, while the larger amplitude mode, F3, shows no obvious harmonics.  More specifically, R$_c$ for 2F4 is significantly larger than 2F1 (the harmonic of an $\ell$=1 mode). However, 2F4's R$_c$ value is closer to what we would expect for an $\ell$=2 harmonic \citep{wu01}. Analysis of the combination modes from the VLT data set is somewhat suspect since they all have low amplitudes and are easily influenced by noise.  Plus, relative amplitudes of combination modes are influenced by azimuthal order. With such a short data set we cannot measure the azimuthal splitting \citep{kleinman98} nor its effects on the combination amplitudes.  While the combinations enforce the idea of higher $\ell$ for F4, they leave some question to its actual identity.

Part of the ambiguity in the identification of F4 may be caused by our entirely linear treatment of pulsation. \citet{ik01} show that pulsation modes can be distributed across the star in a manner significantly different from a simple spherical harmonic.  From their simulations, they find that an $\ell$=3 with large amplitudes can have chromatic amplitudes more like $\ell$=1 or 4 than $\ell$=3. Higher $\ell$ modes must have larger amplitudes in order to be observed, and are more likely to suffer from these nonlinear effects. This could be the reason we observe line shape variations similar to $\ell$=4 modes, but not $\ell$=3; nonlinear mixing distorts the mode identification. In addition to our F4, these nonlinearities may be responsible for the unusual chromatic amplitudes seen on G~117-B15A by \citet{kotak04}. Luckily, knowing that the mode is high $\ell$ is enough to improve asteroseismological modeling. Given these considerations, we cautiously denote F4 as $\ell=$4 or 3. 

Because F4 lies near to the combination of F1 and F2, it is worth considering the possibility that F4 is a naturally damped mode excited by resonance \citep{dz82,gw4}.  However the three-mode coupling angular selection rules show that because F1 and F2 are both $\ell$=1, F4 could only be excited by this mechanism if it were $\ell$=2 \citep{gw4}. Given our measurements of the line shape variations, resonance is unlikely in this instance. 

In Table~2 we give a summary of all mode identifications from time series spectroscopy of G~29-38.  We have eleven identifications, six from the Keck data made by C00, one from the Keck data inferred by \citet{kotak02a}, and four new identification (plus two confirmations) from our VLT data. Note that we have identified modes to be the same if their frequencies matched to within $\sim\!10\,\mu$Hz, which is roughly the rotational splitting inferred by \citet{kleinman98}.

\begin{table*}

\caption{Mode identifications for G~29-38 using time series
spectroscopy. VLT refers to those modes identified in this paper. C00 referes
to \citet{c00} and K02 referes to \citet{kotak02a}. See \S\ref{s:ewd} for 
more details on the 353s mode.}
\begin{minipage}{125pt}
\begin{tabular}{lll}
period & source & $\ell$\\
\hline
\hline
937 & VLT	&1\\
920 & K02   &2\\
835 & VLT	& 1\\
815 & C00,VLT & 1\\
776 & C00	& 2\\
681 & VLT	& 2\\
655 & C00	& 1\\
614 & C00,VLT & 1\\
431 & C00	& 1\\
353 & VLT	&4 or 3\\
284 & C00	& 1\\
\end{tabular}
\end{minipage}
\end{table*}

\section{Conclusions}
Time series spectroscopy has proven to be a useful tool to measure white dwarf pulsation. For G~29-38, it has revealed the physical motion of the pulsation through Doppler shifts and $\ell$ through the variations of line shape. The three sets of spectroscopy on G~29-38, the VLT time series presented here, the Keck time series (VK00 and C00), and the high-resolution Keck observations \citep{thompson03}, have proven that this technique gives consistent results and allows us to measure aspects of the star's pulsation that cannot be obtained through photometry. 

The convective driving theory \citep{gw1,bhill83} is now the favored theory to produce pulsations on DAVs. The velocity modes that are measured here, in VK00 and in \citet{thompson03} show amplitudes as predicted by \citet{gw2}. However, we still lack the appropriate data to test the predictions at higher frequencies. $R_v$ should increase at shorter periods, but G~29-38 sadly does not tend to drive pulsation modes shorter than 400~s as strongly as it does 800~s modes, making it hard to measure the trend without extensive time series spectroscopy. Further spectroscopic observations are also warranted because of the velocity combinations \citep[our F1+F2 and][]{thompson03} not anticipated by any theory. Further observations will determine the general properties of these velocity combinations whether they represent nonlinearities in the velocity curve or are independent pulsations possibly driven by resonance.

The major achievement of time series spectroscopy is the measurement of $\ell$, necessary for asteroseismology. This technique promises to reveal internal properties such as the hydrogen layer mass, helium layer mass, temperature, mass etc. \citep[see for example][]{bradley98,castanheira08}. Modeling white dwarf interiors requires knowledge of $\ell$, otherwise more than one predicted mode could easily fit each observed period, and more than one solution exists for the star. Without any other information, $\ell$=1 is generally assumed for all modes, except for those not easily fit by $\ell$=1.  As we have shown for G~29-38, that assumption is frequently incorrect; 4 of the 11 measured modes are $\ell$$\ne$1.  Past attempts to model G~29-38 resulted in an abundance of possible solutions that could fit the complex set of pulsations \citep{kleinman95}. Now that we have constrained $\ell$ for G~29-28, asteroseismic modeling is more likely to succeed.  Though beyond the scope of this paper, by combining known photometry \citep{kleinman98} with the $\ell$ values presented here, we expect that asteroseismological modeling will finally be able to determine the internal structure of the notorious G~29-38.

\section*{Acknowledgments} %%% Text of acknowledgments runs on after this command.
S. E. Thompson thanks the AAS through the small research grant program and the Crystal Trust for their contribution to the completion of this work. We also thank Dr. Celeste Yeates for help analyzing the combination frequencies.

\bibliography{wdtss}
\bibliographystyle{aj}
%\begin{thebibliography}{wdtss}
%\end{thebibliography}

\end{document}